\documentclass[useAMS,usenatbib]{mn2e}

\usepackage{graphics,epsfig}
\usepackage{graphicx}
\usepackage{amssymb}

\title[\HI\ in XMD galaxies]{\HI\ and star formation in the most metal-deficient galaxies}
\author[Ekta, Jayaram N. Chengalur, S. A. Pustilnik]
{Ekta,$^1$\thanks{ekta@ncra.tifr.res.in,chengalu@ncra.tifr.res.in,sap@sao.ru}
Jayaram N. Chengalur,$^1$\footnotemark[1] 
Simon A. Pustilnik$^2$\footnotemark[1]\\
$^1$ National Centre for Radio Astrophysics, Post Bag 3, Ganeshkhind, Pune 411 007, India\\
$^2$ Special Astrophysical Observatory of RAS, Nizhnij Arkhyz, Karachai-Circasia 369167, Russia}

\DeclareRobustCommand{\ion}[2]{%
\relax\ifmmode
\ifx\testbx\f@series
{\mathbf{#1\,\mathsc{#2}}}\else
{\mathrm{#1\,\mathsc{#2}}}\fi
\else\textup{#1\,{\mdseries\textsc{#2}}}%
\fi}

\newcommand{\kms}{km~s$^{-1}$}
\newcommand{\HI}{\ion{H}{i}}
\newcommand{\HII}{\ion{H}{ii}}
\newcommand{\atoms}{atoms~cm$^{-2}$}

\begin{document}
 
\label{firstpage}

\date{Accepted 2008 . Received 2007 }

\pagerange{\pageref{firstpage}--\pageref{lastpage}} \pubyear{2008}

\maketitle

\begin{abstract}

  We present Giant Metrewave Radio Telescope (GMRT) observations
for three (viz., DDO~68, SDSS~J2104--0035 and UGC~772) of the six most
metal-deficient actively star-forming galaxies known. Although there is a debate as to   
whether these galaxies are undergoing their first
episode of star formation or not, they are `young' in the sense that 
their ISM is chemically unevolved. In this regard, they are the nearest 
equivalents of young galaxies in the early Universe.

All three galaxies, that we have observed, have irregular \HI\ morphologies and 
kinematics, which we interpret as either due to tidal interaction with 
neighbouring galaxies, or the consequences of a recent merger. The remaining 
three of the six most metal-deficient galaxies are also known to have highly 
disturbed \HI\ distributions and are interacting. It is 
interesting because these galaxies were chosen solely on the basis of their 
metallicity and not for any particular signs of interaction. In this sense (i.e.,  
their gas has not yet had time to settle into a regular disc), one could regard 
these extremely metal deficient (XMD) galaxies as `young'. The current star formation 
episode is likely to have been triggered by interaction/merger. It is also possible that 
the tidal interaction has lead to enhanced mixing with metal-poor gas in outer disc, 
and hence to a low gas-phase metallicity in the central star-forming regions.

We also try to determine the threshold gas-density for star-formation
in our sample of galaxies, and find that in general these galaxies do 
not show a one-to-one correspondence between regions of high \HI\ column 
density and regions with current star formation. However, to the extent 
that one can define a threshold density, its value ($\sim$10$^{21}$~atoms~cm$^{-2}$) 
is similar to that in galaxies with much higher metallicity. The highest column 
densities that we detect in regions far outside star-forming regions (i.e., a lower 
limit to the star-formation threshold) are $\sim2\times$10$^{21}$~atoms~cm$^{-2}$.

\end{abstract}

\begin{keywords}
galaxies: dwarf -- galaxies: individual: 
DDO~68 (UGC~5340) -- galaxies: individual: SDSS~J2104--0035 -- galaxies: individual: UGC~772 
-- galaxies: kinematics and dynamics -- radio lines: galaxies
\end{keywords}

\section{INTRODUCTION}
\label{sec:intro}

Blue Compact Galaxies (BCGs) were first identified by \cite{zwicky1965} as
unusual members of the extragalactic zoo. A subset of them were later
recognized to be nearby low-mass galaxies. \cite{sargent1970} found that
the giant \HII\ regions in BCGs have low metal-content and blue colours, typical
of young ($<$10 Myr) stellar populations. As luck would have it, the first
well studied BCGs, viz., I~Zw~18 and 
II~Zw~40 turned out to be fairly unrepresentative of the BCG class. Even after
30 years of study of BCGs, I~Zw~18 continued to hold the record for the lowest 
metallicity, while II~Zw~40 is a galaxy pair in an advanced state of merger, 
which in general is not characteristic of BCGs. Partly because of the extreme 
properties of these two BCGs, and also because of the other unusual properties
of the BCG class as a whole, it was hypothesized that some BCGs, including
I~Zw~18, may be undergoing their first burst of star formation. However,
this was quickly recognized as being unlikely, since it was found that the
outer parts of
many BCGs have red colours, typical of a normal old stellar population. 

\begin{figure*}
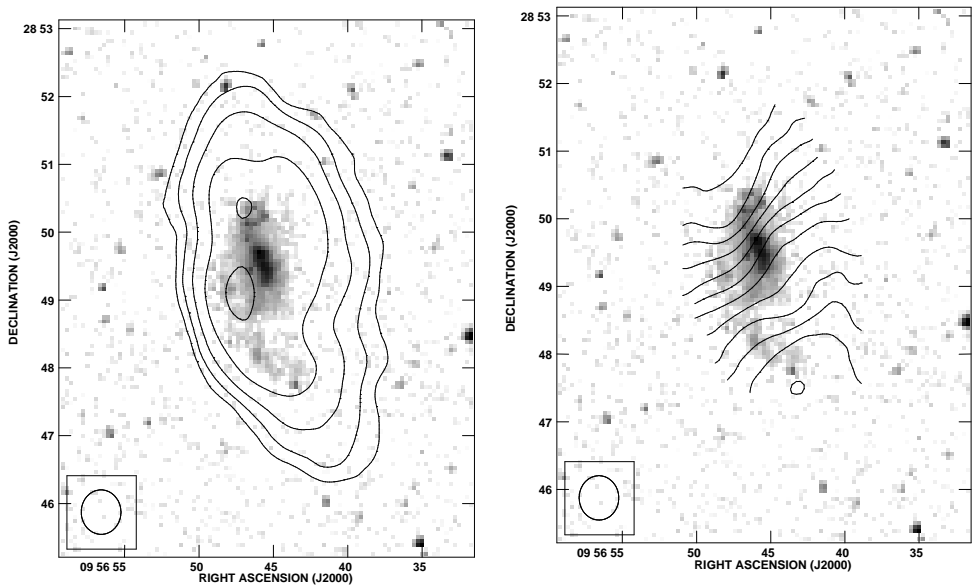

\includegraphics[width=6.5cm]{lowresddo68HI.eps}
\includegraphics[width=6.5cm]{ddo68lowresvel.eps}
\caption{(a.) The integrated \HI\ emission map of DDO~68 (in contours), at
  an angular resolution of $\sim$39~$\times$~35~arcsec$^{2}$, is overlaid
  on $B$-band $DSS$-II optical image (grey-scale). The contours are at
  \HI\ column densities of 7.8, 18.3, 42.8, 99.0 and 233.9
~$\times$~10$^{19}$~\atoms. The grey-scale is in arbitrary units. (b.)
  The \HI\ velocity field at same resolution (contours) overlaid on same
  grey-scale image as in (a). The contours are from 462 to 541 \kms,
  (from south to north) placed at regular intervals of 6.6 \kms.}
\label{fig:lowresddo68HI}
\end{figure*}

None the less the idea that a small fraction of BCGs, particularly those with 
very low metallicity (i.e., the extremely Metal-Deficient (XMD) galaxies, with 
12~+~log(O/H)~$<$~7.65, or Z$<$Z$_{\odot}$/10; see \cite{kunth2000} for a review), 
could be undergoing their first burst of star formation remained popular. In particular, 
deep photometry of the unresolved stellar population in outer parts of several XMD BCGs shows no evidence for an old stellar population, 
e.g., I~Zw~18 (\cite{papaderos2002}), SBS~0335--052~E, W (\cite{papaderos1998}; \cite{pustilnik2004}, but see also \cite{ostlink2001}), 
Tol~65 (\cite{papaderos1999}) and DDO~68 (Pustilnik, Kniazev \& Pramskij 2005;  Pustilnik, Tepliakova \& Kniazev 2007). However, resolved CMD are 
essential to convincingly establish that the current episode of star formation is the first one. The case of I~Zw~18 has been particularly tortuous. From 
$HST$ WFPC2 (\cite{aloisi99}) and NICMOS (\cite{ostlin2000}) observations of this galaxy, it was argued that I~Zw~18 contains AGB stars older than 1~Gyr 
and is hence not undergoing its first episode of star formation. However using deeper HST ACS observations, \cite{izotov2004} concluded that I~Zw~18 
lacked RGB stars and was a bona fide young galaxy. Reanalysis of the same data by other groups indicated that RGB stars were, in fact, 
present (\cite{momany2005}; \cite{tosi2006}). More recently, \cite{aloisi2007} analysed deeper {\it HST} images of I~Zw~18 and convincingly 
established the presence of an RGB population, and hence that I~Zw~18 has been forming stars for $\gtrsim$~2~Gyr.
One more XMD BCG, SBS~1415+437, which was initially claimed to be a
candidate young galaxy (\cite{thuan1999}; but see also \cite{guseva2003}, who reached a significantly more limited conclusion: viz., its low surface 
brightness (LSB) component luminosity-weighted age is probably not greater than 1--2~Gyr),
was also found on the basis of {\it HST} based colour-magnitude diagram (CMD) analysis, to have an RGB population with age $\gtrsim$1.3~Gyr 
(\cite{aloisi2005}). Similar deep CMD data on the above-mentioned candidate young galaxies are needed to rigorously test for the presence of an older 
stellar population.

\begin{table*}
\caption{Parameters of the GMRT observations}
\label{tab:obspar}
\begin{tabular}{llll}
\hline
     & DDO~68 & J~2104--0035 & UGC~772\\
\hline
Date of observations       & 2005 November 19, 20 & 2006 June 07 & 2006 November 25\\
Field center R.A.(2000)    & 09$^{h}$56$^{m}$45.7$^{s}$ & 21$^{h}$04$^{m}$55.3$^{s}$ & 01$^{h}$13$^{m}$40.4$^{s}$ \\
Field center Dec.(2000)    & 28$^{o}$49$^{'}$35.0$^{"}$ & --00$^{o}$35$^{'}$21$^{"}$ & 00$^{o}$52$^{'}$39.0$^{"}$  \\
Central Velocity (\kms)    & 503  & 1401 & 1157\\
Time on-source  (h)        & $\sim$10 & $\sim$6 & $\sim$7.5 \\
Number of channels         & 128  & 128 & 128 \\
Channel separation (\kms)  & $\sim$3.3 & $\sim$3.3 & $\sim$3.3 \\
Flux Calibrators           & 3C48, 3C286 & 3C286 & 3C48\\
Phase Calibrators           & 0824+185, 1111+119 & 2130+050 & 0519+001, 0204--170\\
Resolution (arcsec$^{2}$) (rms (mJy~Bm$^{-1}$)) & 39~$\times$~35 (1.6) & 29~$\times$~25 (1.7) & 29~$\times$~25 (1.7) \\
                                & 27~$\times$~22 (1.1) & 17~$\times$~11 (1.4) & 34~$\times$~26 (1.8) \\
                                & 12~$\times$~10 (1.0) & 10~$\times$~7  (1.2) & 15~$\times$~10 (1.2) \\
                                & 8~$\times$~7   (0.9) & 7~$\times$~5   (1.0) & 8~$\times$~7   (1.1) \\
\hline
\end{tabular}
\end {table*}

However, regardless of whether XMD galaxies are undergoing their first 
episode of star formation, or not, the fact remains that they represent the 
most metal-poor sites of ongoing star formation. In this sense, they can 
still be expected to provide useful insight into the processes that are 
relevant in galaxies at high redshift. For example, a comparison of the
relation between the gas distribution and star formation in these galaxies
with the observed relation in high-metallicity galaxies could help 
determine what, if any, systematic differences there are between 
star formation in high- and low-metallicity gas. In our continuing series
of studies of \HI\ in XMD galaxies
(\cite{chengalur2006}; Ekta, Chengalur \& Pustilnik 2006; \cite{pustilnik2007}),
we present Giant Metrewave Radio Telescope (GMRT) \HI\ 21~cm observations 
of three of the six most metal-poor BCGs known. Global properties of
the sample galaxies are briefly discussed in Section~\ref{sec:sample}, and the
observations are described in Section~\ref{sec:obs}. The morphology and
kinematics of gas in these galaxies is in itself interesting, and is 
discussed in Section~\ref{sec:HI}. In Section~\ref{sec:SF},
we use optical broad-band and/or H$\alpha$ images (taken from the published
literature) as tracers of  star formation (SF), to try to determine
the relation between SF and the gas surface-density. In
Section~\ref{sec:dis}, we combine our results with those already obtained
for other XMD galaxies, and compare them with those obtained from
similar studies of more metal-rich galaxies. Finally, in Section~\ref{sec:summ}, 
we summarise our results.  

\section{THE SAMPLE} 
\label{sec:sample}

   Fresh GMRT observations were obtained for three of the six most metal-deficient actively star-forming galaxies known, viz., DDO~68, UGC~772 and 
SDSS~J2104--0035. Their optical appearance is dominated by the current starburst which is responsible for their blue colours.
DDO~68 (UGC~5340, M$_B$ $\sim$--14.54, discussed below) was identified as an XMD galaxy as part of a survey for dwarfs in the region of a 
nearby Lynx-Cancer minivoid (Pustilnik et al. 2005), while the remaining two galaxies were identified as XMD objects from a search through the SDSS DR4 
data for metal-poor emission-line galaxies (Izotov et al. 2006a).

\cite{pustilnik2005} found that the oxygen abundance of DDO~68 is 
12~+~log(O/H)~$\sim$7.21~$\pm$~0.03, which is in reasonable agreement with (albeit slightly higher than) that measured by \cite{izotov2007} 
(weighted-average~$\sim$7.14~$\pm$~0.03). 
It is the nearest of the 6 lowest
 metallicity XMD galaxies, with an estimated distance of 6.5~Mpc 
  (Pustilnik et al. 2005).
  \footnote[1]{The peculiar velocity flow in the vicinity of the Lynx-Cancer
  void makes distance to DDO~68 uncertain (\cite{Tully2007}).} The same
  authors present $V$, $R$ 
  and H$\alpha$ images for DDO~68, and draw attention to its peculiar optical
  morphology, which consists of central elongated region with a long curved
  tail extending to the south.
The $B$ magnitude given in \cite{pustilnik2005} is m$_{B}$~=~14.60, which
  corresponds to M$_{B}$~=~--14.54, at distance of 6.5~Mpc.

\cite{izotov2006} measured oxygen abundances of
12~+~log(O/H)~=~7.17~$\pm$~0.09 (updated to 7.24~$\pm$~0.05 by
\cite{izotov2007}) and 7.26~$\pm$~0.03 for UGC~772  and SDSS~J2104--0035 (J2104--0035 in the rest of paper, for brevity), 
respectively. It should be noted that for UGC~772 oxygen abundance of only one 
of its several \HII\ regions has been determined using the direct method.

\begin{figure}
\includegraphics[width=8.65cm]{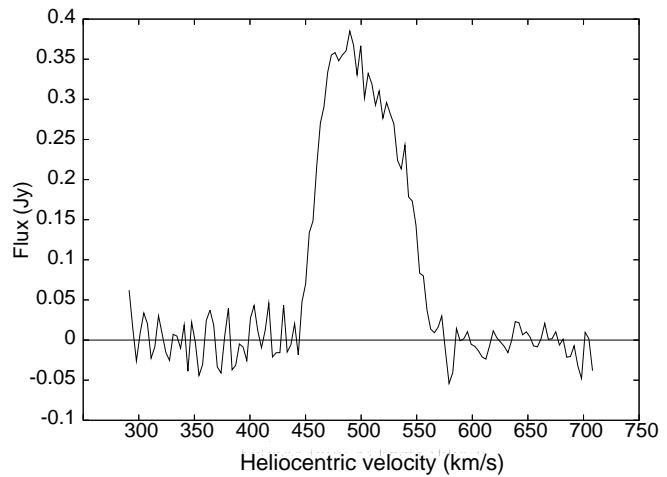}
\caption{The integrated \HI\ spectrum of DDO~68 made from spectral cube at 
	 angular resolution of $\sim$39~$\times$~35~arcsec$^{2}$.}
\label{fig:specddo68}
\end{figure}

UGC~772 is part of a loose group of galaxies, dominated by a SAB(s)m galaxy,
NGC~428. It has a radial velocity of 1157~\kms\ (Smoker, Davies \& Axon 1996), which
corresponds to a Hubble-flow
distance of $\sim$16~Mpc. At this distance, an apparent blue magnitude of
16.28 (Smoker et al. 1996) corresponds to M$_{B} \sim$--14.8.

J2104--0035 has a radial velocity of 1401~\kms, which for
Hubble constant of 70~\kms\~Mpc$^{-1}$
corresponds to a distance of 20~Mpc. In section~\ref{ssec:j2104hi} below, we
estimate its absolute blue magnitude to be $M_B \sim$--13.7. 

\section{OBSERVATIONS AND DATA REDUCTION}
\label{sec:obs}

The details of the GMRT observations are given in Table~\ref{tab:obspar}.
All the data were analysed using the {\footnotesize AIPS} software package.
The major steps in the data analysis were (1) flagging of bad data
points, (2) flux and phase calibrations, (3) bandpass calibration,
(4) subtraction of continuum emission using  {\footnotesize UVSUB},
(5) combining visibility files (in the case of multiple observing runs)
using {\footnotesize DBCON}, (6) imaging and cleaning using the task
{\footnotesize IMAGR}, (7) subtraction of any residual continuum using
{\footnotesize IMLIN}. The imaging and cleaning process was done at a variety
of resolutions, see Table~\ref{tab:obspar} for details.
Note, that since the GMRT has a hybrid configuration (\cite{swarup1991}),
images at resolutions ranging from $\sim$40 to 4~arcsec can be
made from data from a single observing run. Integrated \HI\ 21~cm emission
and velocity fields were
generated from the data cubes using the task {\footnotesize MOMNT}.
J2104--0035 was also observed using the Multibeam receiver of the Parkes
radio telescope.

\section{\HI\ MORPHOLOGY AND KINEMATICS}
\label{sec:HI}

\subsection{DDO~68}
\label{ssec:ddo68HI}

 DDO~68 has been previously observed at the Westerbork Synthesis Radio
Telescope (WSRT) by \cite{stil}. Their maps (which have a relatively high
resolution of $\sim$13.5~arcsec), by and large, trace only the peaks in
the \HI\ column density distribution and resolve out emission from
the more extended, low-column-density gas seen in the low-resolution GMRT
images (e.g.,
Fig.~\ref{fig:lowresddo68HI}). As can be seen in the figure, the \HI\
emission is significantly more extended than the optical emission.
\HI\ emission corresponding to the optical tail to the
south can also be clearly seen. Finally, the \HI\ distribution cuts off
quite sharply on the eastern side, while it falls off much more gradually
to the west. At the same resolution, the velocity field
(Fig.~\ref{fig:lowresddo68HI}) is, to zeroth order, consistent with that
expected from a rotating gas-disc, however, deviations from axi-symmetry can
also be seen. For example, the contours in the northern half of the galaxy
are more open than those in the southern half. The \HI\ spectrum of DDO~68
(obtained from the spectral cube at this same angular resolution)
is shown in Fig.~\ref{fig:specddo68}. The total \HI\ flux that we measure
from this spectrum is 28.9~$ \pm$~3~Jy~\kms, which matches, within the error
bars, with the value of 26.1~Jy~\kms\ measured by \cite{stil}. For our assumed
distance of 6.5~Mpc, the \HI\ mass is 2.9~$\times$~10$^{8}$~M$_{\odot}$.
Like most other XMDs, DDO~68 is gas-rich -- M$_{HI}$/L$_{B}$ $\sim$2.9.

\begin{figure}
\includegraphics[width=8.7cm]{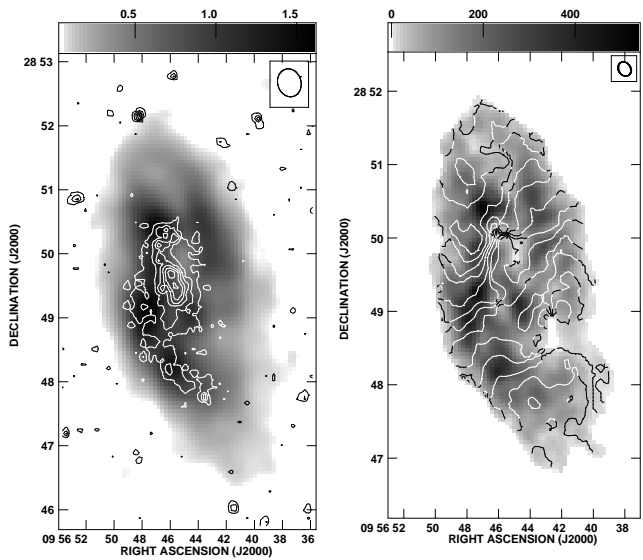}
\caption{{\bf (Left)}The integrated \HI\ emission map of DDO~68, at an angular resolution
  of $\sim$27~$\times$~22~arcsec$^{2}$, in grey-scale, over a range of
  1.0~--~33~$\times$~10$^{20}$~\atoms.
  A $B$-band $DSS$-II optical image is overlaid, in contours, in arbitrary
  units. {\bf (Right)} The \HI\ velocity contours of DDO~68 at an angular resolution of 
$\sim$12\arcsec~$\times$~10\arcsec\ overlaid on integrated \HI\ emission
map (grey-scale)
at same resolution. The contour levels are from 458 to 543.8~\kms\ and 
are regularly spaced at 6.6~\kms. The grey-scale is over the range
of 0 to 534~Jy~Beam$^{-1}$~ms$^{-1}$
(N$_{HI}$~$\sim$5.4~$\times$~10$^{21}$~\atoms.)}
\label{fig:25arcddo68}
\end{figure}

In Fig.~\ref{fig:25arcddo68} we show a higher-resolution
($\sim$27~$\times$~22~arcsec$^{2}$) image of the integrated \HI\ emission
(grey-scale) with the $B$-band $DSS$-{\sc II} optical image overlaid as
contours. At this resolution one can see that the galaxy has two massive
\HI\ arms, North-East-South (NES) and South-West-North (SWN), which wind around
the main stellar body. The southern part of NES arm is coincident with the optical tail, while 
SWN arm has no obvious optical counterpart. Since it is highly unusual for a dwarf galaxy to have spiral arms, we interpret these 
structures (see below) to be tidal tails in a putative merger remnant.  
In addition to these arms/tails, there is
\HI\ emission that roughly coincides with the central stellar body, but its
western part corresponds to regions with very low \HI\ column density.

\begin{table}
\caption{Velocity widths in DDO~68}
\label{tab:dispersion}
\begin{tabular}{llll}
\hline
Region & \cite{stilthesis} & GMRT (22~arcsec) & GMRT (12~arcsec)  \\
           & (\kms)             & (\kms)            & (\kms)            \\ 
\hline
1     &    16.7 &  12.7 & 10.5 \\
2     &    18.3 &  17.0 & 14.4 \\
3     &    14.4 &  10.4 & 9.6  \\ 
\hline
\end{tabular}
\end {table}

\begin{figure}
\includegraphics[width=8.9cm]{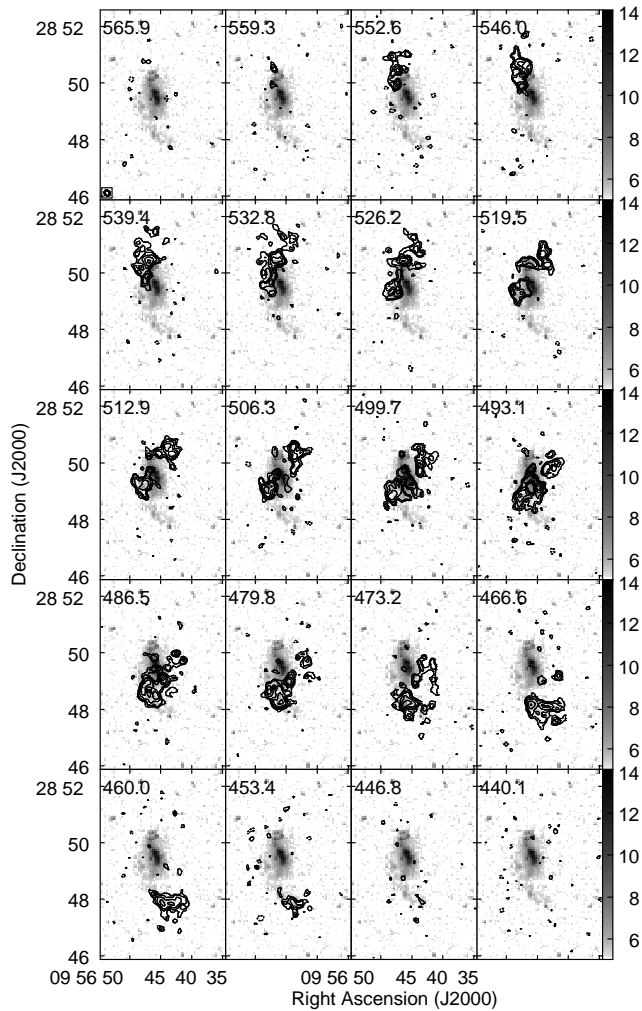}
\caption{The \HI\ channel maps of DDO~68, at an angular resolution of 
$\sim$12~$\times$~10~arcsec$^{2}$, overlaid on a $B$-band $DSS$-{\small II}
optical image, in grey-scale.
The contour levels are at -3, 3, 4.2, 6.0, 8.5 and 12.0 times the rms noise
in single channel, which is 1.1~mJy~beam$^{-1}$. Every alternate channel is
shown and corresponding velocity, in units of \kms, is labelled.}
\label{fig:11arcddo68chan}
\end{figure}

To better understand the connections between the optical and \HI\ structures,
it is instructive to look at the \HI\ channel-maps.
Fig.~\ref{fig:11arcddo68chan} shows channel maps, at resolution of
$\sim$12~$\times$10~arcsec$^{2}$, overlaid on a grey-scale representation
of the $B$-band $DSS$-{\sc II} optical image. One can see that the \HI\
emission that is coincident with the 'southern' optical tail actually starts
at the northern end of the galaxy, at velocities around 553~\kms. As the
velocity decreases, the emission from this NES arm moves southward
along the eastern side of the main body of optical emission, and then
finally forms the long arcing tail to the south. In the lowest velocity
channel maps, one can see that the \HI\ emission extends to much larger
galacto-centric distances than the corresponding stellar emission. In the
channel maps one can also discern the two other \HI\ features seen in
Fig.~\ref{fig:25arcddo68}, viz., the \HI\ associated with the central body
of the galaxy and the SWN \HI\ arm.
Note, however, the overlap of \HI\ is largely on the eastern side of 
optical galaxy -- the western part of the central stellar 
emission has very little gas associated with it. The SWN arm starts
at the southern end of the galaxy (466~\kms), and moves northwards as the
velocity increases. From the high-resolution velocity field
(Fig.~\ref{fig:25arcddo68}), one can see that there is reasonable continuity
in the velocity field as one goes along the arms, but that there are velocity
discontinuities between the arms and the central body.

 We obtained maximum rotational velocities of 55 and 48~\kms at 144 and 168~arcsecs, respectively, 
while trying to derive rotation curves for approaching and receding sides (not shown). 
Given the disturbed, asymmetric nature of the 
velocity field these can only be regarded as indicative rotation curves. The implied dynamical mass (average of the results for the receding and 
approaching sides) is $\sim$~3.2~$\times$~10$^9$~M$_{\odot}$.

The velocity dispersion is an important parameter for systems like DDO~68,  since in the process of merging the gas is expected to be agitated 
(\cite{elmegreen1993}).
\cite{stilthesis} found an enhanced velocity dispersion in some regions of DDO~68. 
We estimated this parameter from our GMRT data via Gaussian fitting in three of four  
regions shown by \cite{stilthesis}, which we could identify. Our data for two beamsizes are shown in
Table~\ref{tab:dispersion} along
with those of \cite{stilthesis}. As seen from this table, GMRT data give significantly smaller widths, 
which in turn decrease with decrease of beamsize. The largest width (Region 2) corresponds to the place 
of overlap between the `tidal' arm and the main body. The velocity width in these regions is hence likely 
to be enhanced because of bulk motions. Thus our data do not support the proposition that there is a significant 
enhancement of velocity dispersion in this galaxy. Our upper limits imply that they, at most, can be enhanced 
by 30--40~per~cent relative to the standard value of 7--8~\kms. This could have important implications for modelling of gas
agitation in the process of strong interactions/mergers. DDO~68 would be an excellent
candidate for comparing numerical models of gas-rich, low-mass mergers with observations.

\subsection{SDSS~J2104--0035}
\label{ssec:j2104hi}

\begin{figure*}
\includegraphics[width=7.5cm,angle=270]{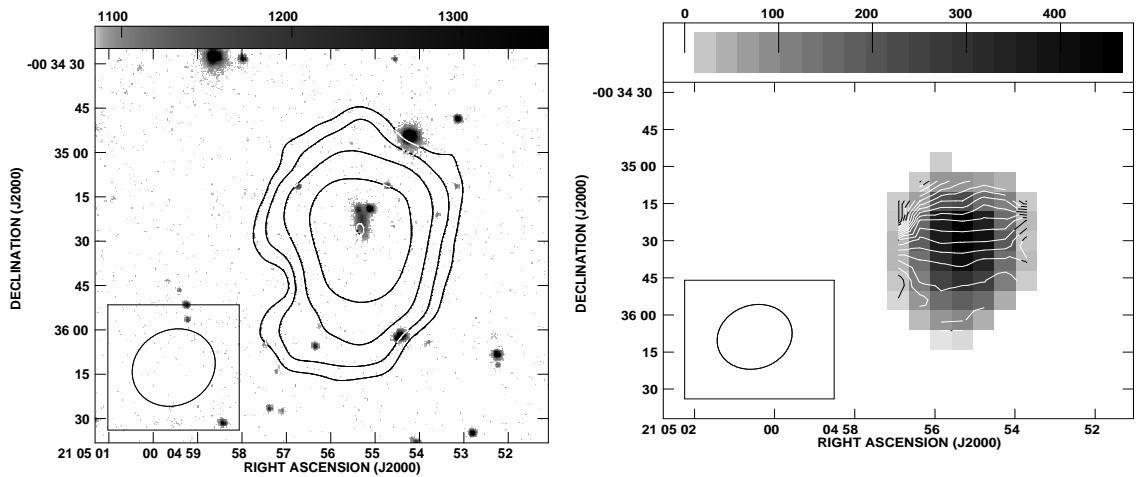}
\caption{{\bf (Left.)} The integrated \HI\ map of J2104--0035, in contours, at an angular
resolution of
$\sim$29~$\times$~25~arcsec$^{2}$, overlaid on a $g$-band SDSS image
(grey-scale in arbitrary units). The contour levels
are at \HI\ column densities of 5.1, 10.6, 22.3, 46.5,
97.3~$\times$~10$^{19}$~\atoms.{\bf (Right.)} The intensity-weighted \HI\ velocity field of J2104--0035, in
contours, overlaid on integrated \HI\ map, in grey-scale, at same angular resolution. The contour levels 
are from 1365 (northern-most) to 1418~\kms\ (southern-most). Grey-scale is at \HI\ column densities from 0 to 455.0~Jy~Beam$^{-1}$~ms$^{-1}$ 
(i.e., N$_{HI}$~$\sim$7.7~$\times$~10$^{20}$~\atoms.)}
\label{fig:27arcj2104hi}
\end{figure*}

  The $g$-band SDSS image of J2104--0035 is shown in
Fig.~\ref{fig:27arcj2104hi}. The optical emission is dominated by two
bright star-forming knots, one at the northern edge of the galaxy
and the other near the centre. \cite{izotov2006} note that its appearance is
similar to that of `cometary' blue compact dwarf galaxies (e.g., Mkn~59, Mkn~71 (\cite{noeske2000})),
which have a bright star-forming region at one end of an elongated stellar
body. This unusual morphology has been suggested to
be a consequence of propagating star formation (\cite{noeske2000}). 
The cometary morphology, combined with the fact that there is a bright star
superposed on the north-western edge of J2104--0035, makes it difficult to
determine the galaxy's surface brightness profile. 
The total magnitude $B_{\rm tot}$ calculated from the total SDSS $u,g,r$
magnitudes is 18.07, which corresponds to $M_B\sim$--13.7,   
at our adopted distance of 20~Mpc. Fitting (adopting an inclination of 70~degrees, as measured from the 
axial ratio of the $g$-band image) only to the southern LSB half 
of the galaxy, (i.e., excluding the star-forming regions) one gets a $B$-band 
scale-length of 2.3~arcsec, and an
(extrapolated) Holmberg radius of 9.8~arcsec ($\sim$0.95~kpc). 

\begin{figure}
\includegraphics[width=5.0cm,angle=270]{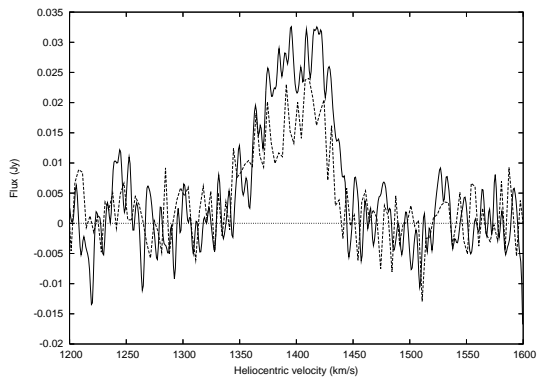}
\caption{The \HI\ spectrum of J2104--0035 obtained from the Parkes telescope
(solid line) and the GMRT (dashed line).}
\label{fig:specj2104}
\end{figure}

     The low (29~$\times$~25~arcsec$^{2}$) resolution \HI\ emission map
(Fig.~\ref{fig:27arcj2104hi}) shows that the \HI\ in J2104--0035 is very
extended compared to the optical emission. At a column density
of 2~$\times$~10$^{19}$ \atoms, the \HI\ extends to $\sim$5.3 times the
Holmberg radius (i.e., an extent comparable to that of DDO~154
(\cite{carignan1998}), though somewhat smaller than that of NGC~3741,
the current record holder for the most extended \HI\ envelope
(\cite{begum2005})). The comparison may not be apt, since unlike the case
for the latter two galaxies, the \HI\ in J2104--0035 is not in a regular
disc. The \HI\ distribution instead looks disturbed with extensions to the
south and east. From the synthetic \HI\ spectrum (Fig.~\ref{fig:specj2104}),
we measure an \HI\ flux of $\sim$1.5~$\pm$~0.15~Jy~\kms. Overlaid on this
spectrum is the spectrum obtained at Parkes, for which the integrated flux
is 2.0~$\pm$~0.2~Jy~\kms. The two measurements do not overlap within the
error bars, suggesting that the galaxy has a diffuse \HI\ component that has
been resolved out at the GMRT. At a distance of $\sim$20.0~Mpc, the Parkes
flux corresponds to a total \HI\ mass of
$\sim$1.9$\pm$0.2$\times$10$^{8}$~M$_{\odot}$, and  a very large
M$_{HI}$ to  L$_{B}$ ratio of $\sim$4.1$\pm$0.4. 
A Gaussian fit to the Parkes spectrum gives a central velocity of 1401~\kms\
and a velocity width of 64~\kms.

\begin{figure}
\includegraphics[width=8.5cm]{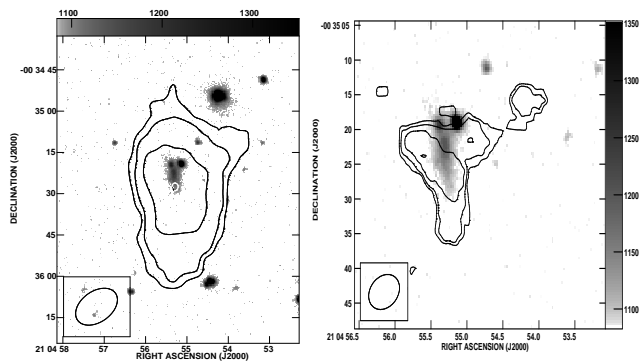}
\caption{{\bf Left.}The integrated \HI\ map of J2104--0035, in contours, at an angular
resolution of
$\sim$17~$\times$~11~arcsec$^{2}$, overlaid on a $g$-band SDSS image
(grey-scale, in arbitrary units). The contour levels are at \HI\ column
densities of 1.0, 2.7, 7.1, 19.2~$\times$~10$^{20}$~\atoms. {\bf Right.} The integrated \HI\ map of J2104--0035 at an angular resolution of 
$\sim$7~$\times$~5~arcsec$^{2}$. The contour levels are at \HI\ column densities of $\sim$6.2, 11.0, 19.3 and 33.8~$\times$~10$^{20}$\atoms. 
The SDSS $g$-band image is shown in grey-scale, in arbitrary units.}
\label{fig:j2104high}
\end{figure}

At higher resolutions (e.g., Fig.~\ref{fig:j2104high}) the \HI\ emission
is a closer match to the optical morphology. The \HI\ emission, however,
remains disturbed-looking with extensions to the south and east.

The intensity-weighted velocity field of J2104--0035, at angular resolution
of $\sim$29~$\times$~25~arcsec$^{2}$, is shown in
Fig.~\ref{fig:27arcj2104hi}. Although there is an overall velocity
gradient from north to south, there is substantial asymmetry between the
northern and southern halves of the galaxy. At the northern tip of the galaxy
the velocity gradient is particularly sharp. Given the large deviations of
the velocity field from that expected from a rotating disc, we do not
attempt to fit a rotation curve to it. Instead we compute an indicative
dynamical mass using,
\begin{equation}
 M_{ind} ~=~2.3~\times~10^{5}~\times~R _{kpc}~\times~V^{2}_{kms^{-1}}~M_{\odot}. 
\label{eqn:mass}
\end{equation}
Assuming an inclination of 70~degrees, a velocity range of
$\sim$41~\kms\ and a radial extent of $\sim$5.0~kpc, gives an indicative
dynamical mass
M$_{ind}$~$\sim$2.4~$\times$~10$^{9}$~M$_{\odot}$.

\subsection{UGC~772}
\label{ssec:ugc772HI}
 
  The optical emission in UGC~772 (with M$_B \sim$16.28 (\cite{smoker1996}))
consists of two well-separated condensations; in the UGC catalog itself, the
system is flagged as being `possibly two dwarfs in contact'. It is part
of a small group of galaxies of which a SAB(s)m galaxy, NGC~428 is the
dominant member (at heliocentric velocity of 1152 \kms\ and with M$_{\rm B}
\sim$--19.2). Smoker et al. (1996) present VLA \HI\ observations of the NGC~428
field. Their maps have low signal-to-noise ratio at position of UGC~772
(which is at the edge of their imaged field), but they note that the galaxy
may have an \HI\ extension to south-east and that the system may represent
the ongoing merger of two smaller dwarfs. They measure a heliocentric
velocity of 1157~\kms (which corresponds to a distance of $\sim$15.1~Mpc, after correcting for Virgo centric infall). At this distance the absolute 
magnitude is M$_{B} \sim$--14.74.

\begin{figure}
\includegraphics[width=5.5cm,angle=270]{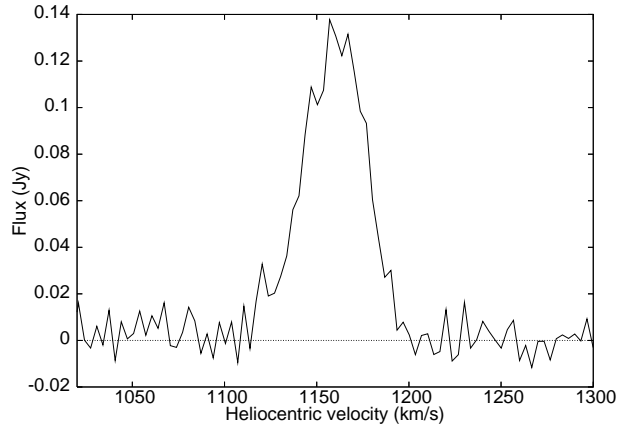}
\caption{The integrated \HI\ spectrum of UGC~772 obtained from data at a 
resolution of $\sim$15$\times$~10~arcsec$^{2}$.}
\label{fig:specugc772}
\end{figure}

The synthetic \HI\ spectrum of UGC~772 (produced from data at resolution of 
$\sim$15$\times$10~arcsec$^{2}$) is shown in Fig.~\ref{fig:specugc772}.
From the spectrum we measure a heliocentric velocity of 1157$\pm$1.7~\kms\
and a total integrated flux of 5.3$\pm$0.4~Jy~\kms, in good agreement with
the single dish measurements of \cite{smoker2000}. From the GMRT
observations, the \HI\ mass is 2.9$\pm$0.2~$\times$~10$^{8}$~M$_{\odot}$, and 
M$_{HI}$/L$_{B}$ $\sim$2.4.

\begin{table}
\caption{Main parameters of the observed galaxies.} 
\label{tab:mainpar}
\begin{tabular}{llll}
\hline
 Parameter    & DDO~68 & J2104--0035 & UGC~772\\
m$_{\rm B}$  & 14.6 & 18.07 & 16.28 \\
A$_{\rm B}^{\rm 1}$  & 0.08 & 0.28  & 0.122 \\
V$_{\rm hel}$ (\kms)  & 502 & 1401 & 1157 \\
D(Mpc) & 6.5 & 20.0 & 15.1 \\
M$_{\rm B}$ & --14.54  & --13.7 & --14.7\\
12+$\log$(O/H) & 7.14~$\pm$~0.03 & 7.26~$\pm$~0.03 & 7.24~$\pm$~0.05 \\
\HI\ flux (Jy~\kms) & 28.9 & 2.0 & 5.3 \\
M$_{\rm HI}$ (10$^{8}$~M$_{\odot}$) & 2.9 & 1.9 & 2.9\\
M$_{\rm gas}$ (10$^{8}$~M$_{\odot}$) & 3.8 & 2.5 & 3.8\\
M$_{\rm HI}$/L$_{\rm B}$ & 2.9 & 4.1 & 2.4 \\
\hline
\end{tabular}
$^{1}$~A$_{B}$ are the values of foreground Galactic extinction(Schlegel, Finkbeiner 
\& Davis 1998) as given in NED.
\end{table}

\begin{figure*}
\includegraphics[width=18.0cm]{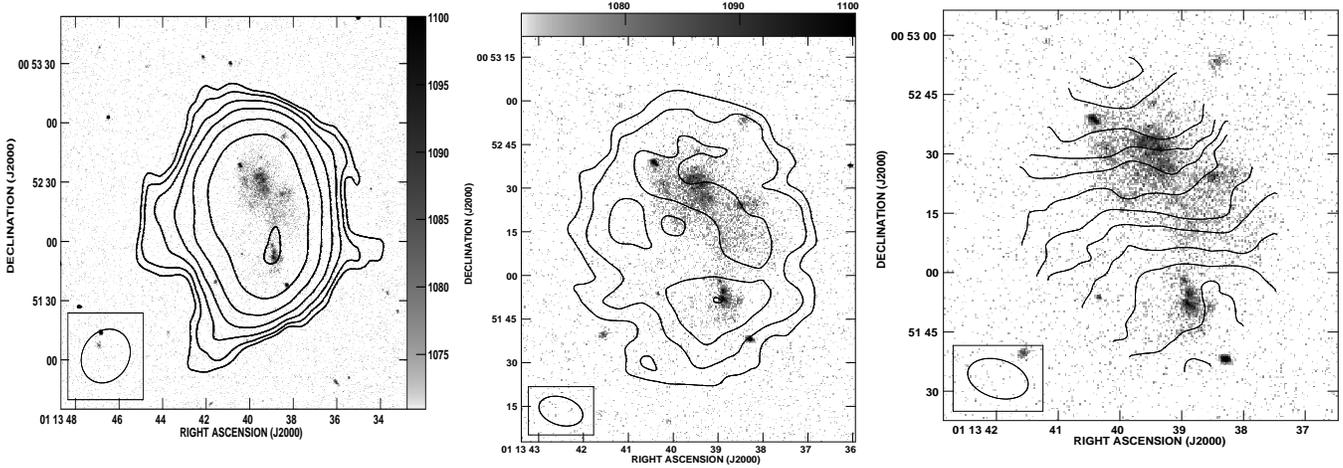}
\caption{{\bf (Left.)} The integrated \HI\ emission map of UGC~772 (in contours), at an
angular resolution of
$\sim$34$\times$26~arcsec$^{2}$. The
contour levels are at \HI\ column densities of 2.7, 5.2, 9.8, 18.4, 34.8,   
65.8 and 124.3~$\times$~10$^{19}$~\atoms. {\bf (Centre.)} Integrated \HI\ emission map of UGC~772, in contours, at a 
resolution of 15$\times$~10~arcsec$^{2}$. The contours 
are at 3.4, 6.6, 12.8, 24.8~$\times$~10$^{20}$~atoms~cm$^{-2}$. {\bf (Right.)} Intensity weighted \HI\ velocity field 
of UGC~772 field, in contours, at an resolution of $\sim$15$\times$~10~arcsec$^{2}$. The contours are equally spaced 
at velocity interval of 3.3~\kms, in the range of 1140--1176~\kms. All contours are overlaid upon SDSS $g$-band optical image in
grey-scale (arbitrary units).}
\label{fig:ugc772}
\end{figure*}

\begin{figure}
\includegraphics[width=8.7cm]{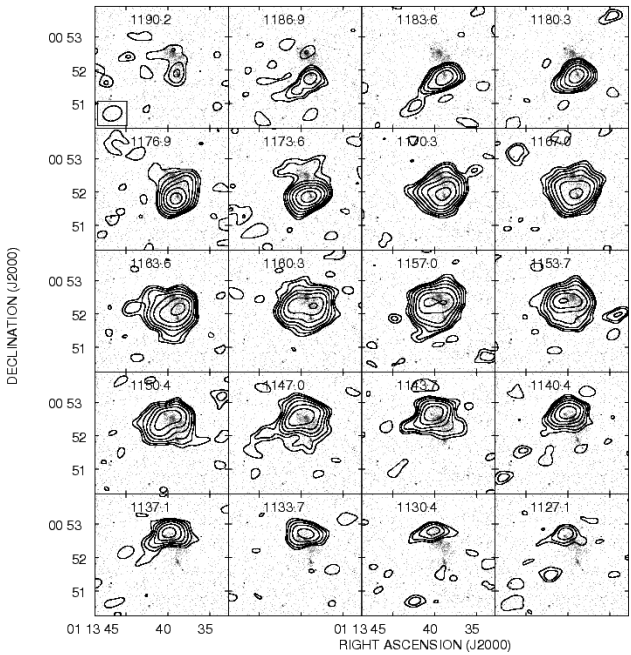}
\caption{The channel maps of UGC~772, at an angular resolution of
$\sim$34$\times$26~arcsec$^{2}$, in contours. Velocity corresponding to
each channel is labelled. Rms noise in single channel is 1.8~mJy~beam$^{-1}$.
The contours are at --2, 2, 2.8, 4.0, 5.7, 8.0, 11.3 and 16.0 times
rms. The SDSS $g$-band image is shown in grey-scale, in arbitrary units.}
\label{fig:ugc772chanmap}
\end{figure}

    Our low-resolution image (Fig.~\ref{fig:ugc772}(Left.)) also shows the
tentatively detected south-east \HI\ extension in the VLA map of Smoker et al. (1996). In the channel maps 
(Fig.~\ref{fig:ugc772chanmap}), these extensions can be seen only over
relatively narrow velocity ranges (viz., 1186.9--1180.3 and
1147.0--1137.1~\kms). At higher resolutions
(e.g., at $\sim$15$\times$10~arcsec$^{2}$, Fig.~\ref{fig:ugc772}(Centre.)),
the \HI\ emission can be seen to have a central `hole' surrounded
by a broken high column density `ring'. The highest \HI\ column
density region of the `ring' overlaps with the bright condensation
to the south. Optical emission is present over the entire north
western part of the `ring'; in contrast, eastern part of
`ring' does not appear to have any associated optical emission.
However, on smoothing the optical emission to a resolution of $\sim$
7~$\times$~7~arcsec$^{2}$, one can see diffuse faint emission coincident with
the eastern part of the `ring'.

The \HI\ velocity map of UGC~772, at a resolution of
$\sim$15$\times$10~arcsec$^{2}$, overlaid with \HI\ integrated map at same resolution, is shown in Fig.~\ref{fig:ugc772}(Right.). The velocity 
field is distorted and the outer parts are suggestive of a warp. We tried to derive rotation curve for approaching and receding sides (not shown). 
Leaving the inclination as a free parameter in the fit did not lead to meaningful results, hence the inclination 
was set to 
a value of 40~degrees, as obtained from the \HI\ morphology and assuming an intrinsic axial ratio of 0.25. 
The maximum rotational velocities are 26 and 29~\kms, at radial distances of 42 and 54~arcsecs, for approaching 
and receding sides, respectively. As for DDO~68, we caution that this rotation 
curve should only be treated as indicative. The derived (average of receding and approaching sides) 
dynamical mass is 8.3~$\times$~10$^8$~M$_{\odot}$.  

\begin{figure*}
\includegraphics[width=17.5cm]{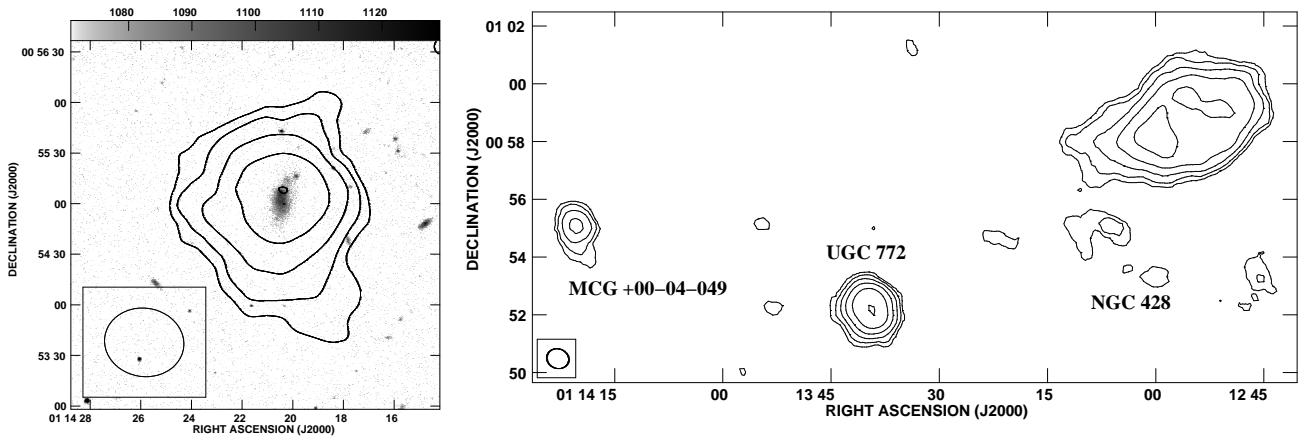}
\caption{{\bf (Left.)} Integrated \HI\ emission map of MCG~+00--04--049, in contours, at a resolution of $\sim$46~$\times$~42~arcsec$^{2}$ 
($\sim$3.2~kpc$^{2}$). 
The contour levels are at \HI\ column densities of $\sim$7.2, 12.3, 21.0, 35.9, 61.3~\atoms. The primary-beam gain-factor at 
position of MCG~+00--04--049 is $\sim$1.7. The $g$-band SDSS image is shown in grey-scale, in arbitrary units. {\bf (Right.)} 
Integrated \HI\ emission map of UGC~772 field at same resolution as in (left.). MCG~+00--04--049 is seen in east, UGC~772 is central, 
while NGC~428 is the westernmost. 
The contour levels are at 65, 123, 234, 443, 843, 1600~Jy/B*ms$^{-1}$. 1~arcsec corresponds to a linear scale of $\sim$73~pc.
}
\label{fig:ugc772field}
\end{figure*}

  A larger-scale map of the field (Fig.~\ref{fig:ugc772field}) shows two
more galaxies, viz., NGC~428 and MCG~+00--04--049 (note that this image is
not corrected for primary-beam attenuation). NGC~428 shows a tail
to the south, which is better seen in the VLA data of Smoker et al. (1996),
in which this galaxy is at the center of the field. This tail, the
south-east extension in \HI\ emission of UGC~772
(Figs.~\ref{fig:ugc772}(Left.),~\ref{fig:ugc772chanmap}) and the south-west
tail-like extension in the outer \HI\ contours of MCG~+00--04--049
(Fig.~\ref{fig:ugc772field}(Left.)) strongly suggest that the galaxies in the group are
tidally interacting.

\section{STAR FORMATION}
\label{sec:SF}

   It has been suggested that star formation in dwarf galaxies occurs only
when the gas column-density crosses a threshold value of $\sim$10$^{21}$~\atoms (\cite{skillman1987}; 
\cite{taylor94}). If this is related to a threshold amount of dust shielding 
required for the production of molecular gas, one might expect that XMD galaxies
would have a higher threshold column-density.

   The H$\alpha$ image of DDO~68 (\cite{pustilnik2005}), shows three separate regions of ongoing 
star formation, one in the central region, the second in the northern ring and the third in a 
ring-like \HII\ region in the southern tail. On comparing the H$\alpha$ map with the \HI\ column 
density at a resolution of $\sim$8~$\times$~7~arcsec$^{2}$ (linear resolution of $\sim$250~pc), we
find that, in general (but not in all cases), the regions of ongoing star formation correspond to 
peaks in the \HI\ column density. However, there are several H$\alpha$ emitting regions for which the gas
surface-density (computed after multiplying the \HI\ face-on column density by a factor of 1.3 to 
account for Helium) is lower (by a factor of $\sim$2) than the nominal threshold density of 
10$^{21}$~\atoms. Further, there are regions where the gas surface-density is substantially 
higher than that the threshold density (e.g., in the tail), but for which there is no
corresponding H$\alpha$ emission.

   In Fig.~\ref{fig:j2104high} we show a $\sim$6.7~$\times$~4.7~arcsec$^{2}$-resolution
($\sim$645~$\times$~459~parsec$^{2}$) map of the \HI\ emission from the
galaxy. Though both \HII\ regions in J2104--0035 are close to high \HI\
column density regions, Fig.~\ref{fig:j2104high} shows that they lie
outside the region of highest \HI\ column density. The gas surface-density at 
the northern \HII\ region is only $\sim$6.0~$\times$~10$^{20}$~\atoms. 
Further, there are \HI\ peaks in western and southern 
\HI\ extensions, at gas densities close to threshold, but with no
corresponding optical emission.

 For UGC~772, we calculated gas surface-densities at locations of the
\HII\ regions, from a map at an angular resolution of
$\sim$8~$\times$7~arcsec$^{2}$, i.e., $\sim$600~$\times$~500~pc$^{2}$.
The peak gas surface-density (computed assuming inclination 
$\sim$ 42~degrees) $\sim$3.1~$\times$~10$^{21}$~\atoms), is associated with the southern
optical condensation. \HII\ regions, marked as 1 and 3 in Fig.~1 of
\cite{izotov2007}, are associated with \HI\ peaks at gas surface-densities
$\sim$1.9 and 1.2~$\times$~10$^{21}$~\atoms, respectively.
On the other hand, the \HII\ region 2 of \cite{izotov2007} lies in the
central \HI\ `hole', with a corresponding gas surface-density  of 
$\sim$5.6~$\times$~10$^{20}$~\atoms. Further, there are regions in 
the eastern part of the \HI\ `ring', which have gas surface-densities 
greater than 2.0~$\times$~10$^{21}$~\atoms, but with no associated 
optical emission.

In summary for these three galaxies, a gas column density of $\sim 10^{21}$~\atoms
does not appear to be either a necessary or sufficient condition for star
formation. Note that in all cases, our spatial resolution is comparable
or better than the $\sim 500$~pc linear resolution used to determine the
threshold density by Skillman (1987).

    \cite{schaye2004} (see also \cite{schaye2007, schayed2007}) suggested that star formation is related to the formation of a cold phase of the interstellar medium, and gives fitting formulae for the threshold column density as a function of the gas fraction, pressure, metallicity,  ionizing radiation flux, etc.. All other factors being equal, the threshold density increases with increasing gas fraction and decreasing metallicity. This would imply, on the face of it, that gas rich XMD galaxies should have a higher threshold for star formation than spirals. However as discussed in detail in \cite{schaye2004}, for determining the star formation threshold in spirals, it is the conditions in the outer portions of the disc 
that are relevant. In these regions, the gas fraction and metallicity are not very different to that in XMD galaxies. If we assume that that in the outer parts of spiral discs, the gas fraction is 1.0, the thermal pressure accounts for half the total pressure, the metallicity is 0.1~times the solar metallicity and the $UV$ radiation intensity is $10^6$cm$^{-2}$s$^{-1}$, the expected threshold density from equation~(23) of \cite{schaye2004} is  log(N$_{HI}$)~$ \sim$20.84. On the other hand, for XMD galaxies, if we take that the gas fraction is $\sim$1, the metallicity is $\sim$0.03 solar, but the ratio of thermal pressure to total pressure and the $UV$ intensity are unchanged, the 
predicted threshold density is only a factor  of $\sim$1.4 larger, i.e., log(N$_{HI}$)~$\sim$21. As such, it appears that a much larger
sample will be required to test for the dependence of the threshold
density on the gas-phase metallicity.

\section{DISCUSSION} 
\label{sec:dis}

In Table~\ref{tab:mainpar}, we summarise the main observed and derived
parameters of the studied galaxies. Their $B$-band luminosities all fall within a factor of 2.5, with the average M$_{\rm B} \sim$--14.3, while 
their M(\HI) 
and M(\HI)/L$_{\rm B}$ ratios differ by less a factor of 1.5 and 1.7, respectively. In this respect, they form a much more homogeneous sample than 
the sample of 22 XMD galaxies studied by Pustilnik \& Martin (2007). Of course a sample of 3 is too small to draw any meaningful statistical conclusion, 
this homogeneity is suggestive of similar evolutionary path-ways for all three galaxies.
 All of the three galaxies that we have observed also have peculiar \HI\ morphology and kinematics. UGC~772 is located in a loose 
group, and has easily identifiable neighbours which could be the cause
of the observed disturbances. DDO~68, on the other hand, is located in a low-density region, in the periphery of the Lynx-Cancer void 
(Pustilnik et al. 2005). The Hubble flow around DDO~68 is very quiet. The radial velocities of all six dwarf galaxies within 500~kpc of it differ by 
less than 40~\kms. The nearest neighbour of DDO~68, 
the dwarf galaxy UGC~5427, is located at a projected distance of $\sim$200~kpc and a radial velocity separation of 5~\kms. UGC~5427 also has a somewhat 
disturbed appearance with a chain of bright knots in a spiral-arm like feature, which makes it tempting to suggest that these two galaxies are tidally 
interacting. However, given its relatively large projected separation, it seems unlikely that tidal interaction with UGC~5427 is the cause of DDO~68's 
disturbed morphology. For example, if we assume that their true spatial separation and relative velocity are 200~kpc and $\sim$50~\kms, respectively, 
then their close encounter could occur $\sim$4~Gyr ago. 
This is large compared to the ages ($\sim$0.01--1~Gyr (Pustilnik et al. 2005; Pustilnik et al. 2007)) of the star-formation episodes in DDO~68.

If interaction with UGC~5427 is not the cause of DDO~68's disturbed morphology, then a remaining possibility is that DDO~68 
represents the late-stage merger of two gas-rich progenitors, in which the gas is now settling down to form a disc. 
In this context, it is interesting to note that 
numerical simulations (e.g., \cite{springel}) show that a disc galaxy can form in 
merger of extremely gas-rich progenitors. Mergers are not very rare among BCGs. E.g., \cite{ostlin2001} also interpret, from the disturbed optical 
morphology and H$\alpha$ kinematics of their sample of BCGs, that they are most likely the results of mergers. Similarly, \cite{pustilnik2001BCG} 
assign $\sim$15~per~cent of their sample BCGs as having a merger-like morphology.

We use a very approximate calculation below to show that 
the ages of the oldest stellar populations in DDO~68 ($\sim$1~Gyr), 
as derived by Pustilnik et al. (2007), match with the time of first encounter between its progenitors. Figure~11 of \cite{springeletal} indicates that 
`short' tails, like those of DDO68, are seen during later phases of merger (in comparison to the extended ones produced during the first encounter). Their 
ages are comparable to the rotation period of merging components and one-third of the time since the first encounter. Assuming that the velocity of tail 
relative to the systemic velocity of DDO~68 is $\sim$50~\kms and its length is $\sim$5~kpc,  age of tail is $\sim$200~Myr, which corresponds to a time 
since the first encounter of $\sim$0.6~Gyr.

  We note, however, that although the morphology and large-scale kinematics of DDO~68 suggest that it could be an ongoing merger, its 
large gas-mass fraction, and the lack of evidence for an enhanced velocity dispersion are in conflict with the predictions of 
numerical models of mergers (Springel et al. (2005) and Elmegreen et al. (1993), respectively). In the Springel et al. (2005) models, 
the gas-fraction of the merger product is (f$_{\rm gas} \sim$0.2--0.3), 
while DDO~68 is gas-rich ($M_{HI}/L_{B} \sim$2.9 (Table~\ref{tab:mainpar}); f$_{gas} \sim$0.95 (Pustilnik et al. 2007)). These differences may be, 
in part, due to the fact that the masses of the putative progenitors of DDO~68 are 
roughly two orders of magnitude smaller than those of the model galaxies in the Elmegreen et al. (1993) and Springel et al. (2005) simulations, and 
in part, 
due to limitations in the star formation and feed back recipes used in the simulations. Given the importance of gas-rich low-mass mergers in the early 
universe, a detailed numerical model of DDO~68 would be particularly interesting.

J2104--0035 is also relatively isolated, and the only other galaxy
within 500~kpc and 500~\kms\ from it is the compact dwarf emission-line galaxy SDSS~J210347.23--004949.7 (M$_{\rm B}$~=~--14.49), which is at a projected 
distance of 22.3~arcmin ($\sim$130~kpc for our assumed distance of 20~Mpc) and has, within cited error bars of $\sim$5~\kms, the same redshift. 
This is the only likely candidate for having caused the observed disturbance in J2104--0035. Alternatively we conjecture that J2104--0035 may be a 
case of an advanced merger.

  It is interesting to note that all three of our sample galaxies have
disturbed morphology and kinematics. In this regard, they closely resemble the three lowest metallicity BCGs, viz., I~Zw~18 and SBS~0035--052~E,~W. 
\cite{pustilnik2001} show that SBS~0035--052 consists of an interacting pair of dwarf galaxies with a common \HI\ envelope, while \cite{zee98IZw} show that 
I~Zw~18 has a complex \HI\ distribution, which they describe as a `fragmenting \HI\ cloud in the early stages of galaxy formation'. Thus, although tidal 
interaction was not specifically selected for when searching for XMD BCGs, all six of the most metal-deficient galaxies  have highly disturbed 
morphologies and kinematics. Indeed this seems to be a common feature for XMD BCGs (e.g., \cite{chengalur95}; \cite{chengalur2006}; Ekta et al. 2006).

Our, and past, results suggest that interactions/mergers could play
an important role in triggering star formation in XMD galaxies. There are two factors that we can speculatively suggest as being responsible for this. The 
first is that intense SF episodes may occur only due to strong tidal disturbances (particularly, if the gas discs are less susceptible to internal 
instabilities, see the discussion below). The second is that tidal interaction could lead to a more efficient
large-scale mixing of the gas. This brings lower-metallicity gas from the  outer parts of the galaxy to its centre, resulting in a lower metallicity in 
the central star-forming regions of the gas distribution. In principle, one could check the validity of this hypothesis by looking for statistical 
differences in the metallicity of BCGs with and without tidally induced star formation, or by detailed modelling of gas-rich mergers (e.g., Bekki 2008).

  Regardless of how the SF is triggered, as far as the relationship between the gas distribution and regions of current SF is concerned, XMD BCGs do not 
appear to be qualitatively different from other late-type dwarf galaxies with low metallicities. For a sample of dwarf irregular galaxies 
(with metallicities mostly in XMD regime), \cite{begum2006} find, much as we discuss above, that although SF generally occurs in regions of relatively 
high \HI\ column density, there is no one-to-one relation between gas at high column density and H$\alpha$ emission. This lack of correspondence could be 
due to several effects, for example, ionization of the gas by $UV$ radiation from \HII\ regions could reduce \HI\ column density in regions with 
strong H$\alpha$ flux. Further, since the massive star formation that the H$\alpha$ emission traces are short-lived, stochasticity in SF would also 
reduce the correlation between the instantaneous observed H$\alpha$ flux and the \HI\ column density. It is also worth noting that, to the extent that 
one can determine a threshold density for SF, there does not seem to be any significant difference between XMD galaxies and the more metal-rich 
dwarfs in the Taylor et al. (1994) sample.

On the other hand, in the model of \cite{schaye2004} a lower limit to
the threshold column density can be derived from the largest column 
densities outside the sites of current SF. In our galaxies
this gas-density (corrected for inclination and with account of He) is
$\sim$2$\times$10$^{21}$~atoms~cm$^{-2}$ for DDO~68 and UGC~772, and
$\sim$1.2--2.0$\times$10$^{21}$~atoms~cm$^{-2}$ for J2104--0035 (as inclination of 
corresponding region is uncertain, gas-density may lie in the given range). For a
proper comparison with more metal-rich galaxies, however, one would
need to compare the column densities at a similar linear resolution.

Finally, we note that the baryonic masses of the XMD galaxies in our
sample are all sufficiently large (greater than several 10$^8$~M$_{\odot}$), such that current star burst is not strong enough  
to cause a significant loss of metals through escape of metal-enriched gas (\cite{fragile2004}). The low ISM metallicity of 
these galaxies could be due to a generally subdued star-formation rate in the past (in case of LSB galaxy progenitor). 
For those XMDs for which no tracers of old 
stellar population have so far been found in outer parts, the option of this interaction inducing the first SF episode remains a possibility. 

\section{Summary}
\label{sec:summ}

\begin{enumerate}
\item All of the three XMD galaxies observed at the GMRT, viz., DDO~68, 
      UGC~772 and SDSS~J2104--0035 show disturbed morphology and velocity fields. 
\item These distortions are suggestive of either an ongoing merging or tidal interaction
      with companions.
\item These results, taken together with existing \HI\ data for I~Zw~18 and 
      SBS~0335--052~E and W, suggest that all six of the most metal-poor 
      actively star-forming galaxies in the local universe are either merging or 
      interacting objects. 
\item  Although SF generally occurs in regions of relatively high 
       \HI\ column density, there is no one-to-one relation between gas 
       at high column density and H$\alpha$ emission. To the extent that 
       one can determine a threshold density for SF, there does not seem 
       to be any significant difference between these XMD galaxies and 
       more metal-rich dwarf galaxies.
\end{enumerate}
~\\
{\bf ACKNOWLEDGMENTS} \\
~\\
We thank the staff of the GMRT who have made these observations possible. The 
GMRT is run by the National Centre for Radio Astrophysics of the Tata Institute 
of Fundamental Research.
Partial support for this work was provided by ILTP grant B-3.13, and through 
the Russian Foundation for Basic research grant 06--02--16617. 
Funding for the Sloan Digital Sky Survey (SDSS) and SDSS-II has been provided by 
the Alfred P. Sloan Foundation, the Participating Institutions, the National Science 
Foundation, the U.S. Department of Energy, the National Aeronautics and Space Administration, 
the Japanese Monbukagakusho, and the Max Planck Society, and the Higher Education Funding 
Council for England. The SDSS Web site is http://www.sdss.org/. The SDSS is managed by the 
Astrophysical Research Consortium (ARC) for the Participating Institutions. The Participating 
Institutions are the American Museum of Natural History, Astrophysical Institute Potsdam, University 
of Basel, University of Cambridge, Case Western Reserve University, The University of Chicago, Drexel 
University, Fermilab, the Institute for Advanced Study, the Japan Participation Group, The Johns Hopkins 
University, the Joint Institute for Nuclear Astrophysics, the Kavli Institute for Particle Astrophysics 
and Cosmology, the Korean Scientist Group, the Chinese Academy of Sciences (LAMOST), Los Alamos National 
Laboratory, the Max-Planck-Institute for Astronomy (MPIA), the Max-Planck-Institute for Astrophysics (MPA), 
New Mexico State University, Ohio State University, University of Pittsburgh, University of Portsmouth, 
Princeton University, the United States Naval Observatory, and the University of Washington.

\label{lastpage}


\begin{thebibliography}{1}
\bibitem[\protect\citeauthoryear{Aloisi, Tosi \& Greggio 1999}{}]{aloisi99}
Aloisi A., Tosi M., Greggio L., 1999, AJ, 118, 302
\bibitem[\protect\citeauthoryear{Aloisi et al. 2005}{}]{aloisi2005} 
Aloisi A., van der Marel R.P., Mack J., Leitherer C., Sirianni M., Tosi M., 2005, ApJ, 631, L45
\bibitem[\protect\citeauthoryear{Aloisi et al.}{2007}]{aloisi2007} 
Aloisi A. et al., 2007, ApJ, 667, L151
\bibitem[\protect\citeauthoryear{Begum, Chengalur \& Karachentsev 2005}{}]{begum2005} 
Begum A., Chengalur J.N., Karachentsev I.D., 2005, A\&A, 433, L1
\bibitem[\protect\citeauthoryear{Begum et al.}{2006}]{begum2006} 
Begum A., Chengalur J.N., Karachentsev I.D., Kaisin S.S., Sharina M.E., 
2006, MNRAS, 365, 1220 
\bibitem[\protect\citeauthoryear{Bekki}{2008}]{bekki2008} 
Bekki K., 2008, MNRAS letters, 388, 10 
\bibitem[\protect\citeauthoryear{Carignan \& Purton 1998}{}]{carignan1998} 
Carignan C., Purton C., 1998, ApJ, 506, 125
\bibitem[\protect\citeauthoryear{Chengalur, Giovanelli \& Haynes 1995}{}]{chengalur95} 
Chengalur J.N., Giovanelli R., Haynes M.P., 1995, AJ, 109, 2415
\bibitem[\protect\citeauthoryear{Chengalur et al. 2006}{}]{chengalur2006} 
Chengalur J.N., Pustilnik S.A., Martin J.-M., Kniazev A.Y., 2006, MNRAS, 371, 1849
\bibitem[\protect\citeauthoryear{Elmegreen, Kaufman \& Thomasson 1993}{}]{elmegreen1993} 
Elmegreen B.G., Kaufman M., Thomasson M., 1993, ApJ, 412, 90
\bibitem[\protect\citeauthoryear{Ekta et al.}{2006}]{ekta2006}
Ekta, Chengalur J.N., Pustilnik S.A., 2006, MNRAS, 372, 853
\bibitem[\protect\citeauthoryear{Fragile, Murray \& Lin 2004}{}]{fragile2004} 
Fragile P.C., Murray S.D., Lin D.N.C. 2004, ApJ, 617, 1077
\bibitem[\protect\citeauthoryear{Guseva et al.}{2003}]{guseva2003} 
Guseva N.G., Papaderos P., Izotov Y.I., Green R.F., Fricke K.J., Thuan T.X., Noeske K.G., 
2003, A\&A, 407, 105 
\bibitem[\protect\citeauthoryear{Izotov \& Thuan}{2004}]{izotov2004} 
Izotov Y.I., Thuan T.X., 2004, ApJ, 616, 768
\bibitem[\protect\citeauthoryear{Izotov \& Thuan}{2007}]{izotov2007} 
Izotov Y.I., Thuan T.X., 2007, ApJ, 665, 1115
\bibitem[\protect\citeauthoryear{Izotov et al.}{2006}]{izotov2006} 
Izotov Y.I., Papaderos P., Guseva N.G., Fricke K.J., Thuan T.X., 2006, A\&A, 454, 137
\bibitem[\protect\citeauthoryear{Kunth \& {\"O}stlin}{2000}]{kunth2000} 
Kunth D., {\"O}stlin G., 2000, A\&ARv, 10, 1
\bibitem[\protect\citeauthoryear{Momany et al. 2005}{}]{momany2005} 
Momany Y., et al., 2005, A\&A, 439, 111 
\bibitem[\protect\citeauthoryear{Noeske et al. 2000}{}]{noeske2000} 
Noeske K.G., Guseva N.G., Fricke K.J., Izotov Y.I., Papaderos P., Thuan T.X., 2000, A\&A, 361, 33
\bibitem[\protect\citeauthoryear{{\"O}stlin 2000}{}]{ostlin2000}
{\"O}stlin G., 2000, ApJ, 535, L99
\bibitem[\protect\citeauthoryear{{\"O}stlin \& Kunth 2001}{}]{ostlink2001} 
{\"O}stlin G., Kunth D., 2001, A\&A, 371, 429
\bibitem[\protect\citeauthoryear{{\"O}stlin et al.}{2001}]{ostlin2001}
{\"O}stlin G., Amram P., Bergvall N., Masegosa J., Boulesteix J., M\'arquez I., 2001, A\&A, 374, 800
\bibitem[\protect\citeauthoryear{Papaderos et al. 1998}{}]{papaderos1998} 
Papaderos P., Izotov Y.I., Fricke K.J., Thuan T.X., Guseva N.G., 1998, A\&A, 338, 43
\bibitem[\protect\citeauthoryear{Papaderos et al. 1999}{}]{papaderos1999} 
Papaderos P., Fricke K.J., Thuan T.X., Izotov Y.I., Niklas H., 1999, A\&A, 352, L57
\bibitem[\protect\citeauthoryear{Papaderos et al. 2002}{}]{papaderos2002} 
Papaderos P., Izotov Y.I., Thuan T.X., Noeske K.G., Fricke K.J., Guseva N.G., Green R.F., 2002, A\&A, 393, 461
\bibitem[\protect\citeauthoryear{Pustilnik et al.}{2001a}]{pustilnik2001} 
Pustilnik S.A., Brinks E., Thuan T.X., Lipovetsky V.A., Izotov Y.I., 2001, AJ, 121, 1413
\bibitem[\protect\citeauthoryear{Pustilnik et al.}{2001b}]{pustilnik2001BCG} 
Pustilnik S.A., Kniazev A.Y., Lipovetsky V.A., Ugryumov A.V., 2001, A\&A, 373, 24
\bibitem[\protect\citeauthoryear{Pustilnik, Kniazev \& Pramskij 2004}{}]{pustilnik2004} 
Pustilnik S.A., Kniazev A.Y., Pramskij A.G., 2004, A\&A, 425, 51
\bibitem[\protect\citeauthoryear{Pustilnik et al.}{2005}]{pustilnik2005} 
Pustilnik S.A., Kniazev A.Y., Pramskij A.G., 2005, A\&A, 443, 91
\bibitem[\protect\citeauthoryear{Pustilnik \& Martin 2007}{}]{pustilnik2007} 
Pustilnik S.A., Martin J.-M., 2007, A\&A, 464, 859
\bibitem[\protect\citeauthoryear{Pustilnik et al. 2008}{}]{pustilniketal2007} 
Pustilnik S.A., Tepliakova A.L., Kniazev A.Y., 2008, Astron. Lett., 31, 457
\bibitem[\protect\citeauthoryear{Sargent \& Searle}{1970}]{sargent1970} 
Sargent W.L.W., Searle L., 1970, ApJ, 162, L155
\bibitem[\protect\citeauthoryear{Schaye}{2004}]{schaye2004} 
Schaye J., 2004, ApJ, 609, 667
\bibitem[\protect\citeauthoryear{Schaye}{2007}]{schaye2007} 
Schaye J., 2007, preprint (astro--ph/0708.3366)
\bibitem[\protect\citeauthoryear{Schaye \& Dalla Vecchia}{2008}]{schayed2007} 
Schaye J., Dalla Vecchia C., 2008, MNRAS, 383, 1210 
\bibitem[\protect\citeauthoryear{Schlegel, Finkbeiner \& Davis 1998}{}]{schlegel}
Schlegel D.J., Finkbeiner D.P., Davis M., 1998, ApJ, 500, 525 
\bibitem[\protect\citeauthoryear{Skillman 1987}{}]{skillman1987} 
Skillman E.D., 1987, in NASA Conf. Publ Vol. 2466, Star Formation in Galaxies, cWashington, p. 263
\bibitem[\protect\citeauthoryear{Smoker et al. 1996}{}]{smoker1996} 
Smoker J.V., Davies R.D., Axon D.J., 1996, MNRAS, 281, 393
\bibitem[\protect\citeauthoryear{Smoker et al.}{2000}]{smoker2000} 
Smoker J.V., Davies R.D., Axon D.J., Hummel E., 2000, A\&A, 361, 19
\bibitem[\protect\citeauthoryear{Springel \& Hernquist 2005}{}]{springel} 
Springel V., Hernquist L., 2005, ApJ, 622, L9
\bibitem[\protect\citeauthoryear{Springel, Di Matteo \& Hernquist}{2005}]{springeletal} 
Springel V., Di Matteo T., Hernquist L., 2005, MNRAS, 361, 776
\bibitem[\protect\citeauthoryear{Stil}{1999}]{stilthesis} Stil J.M., 1999, Ph.D. Thesis, 
Leiden University
\bibitem[\protect\citeauthoryear{Stil \& Israel}{2002}]{stil}
Stil J.M., Israel F.P., 2002, A\&A, 389, 29
\bibitem[\protect\citeauthoryear{Swarup et al. 1991}{}]{swarup1991}
Swarup G., Ananthakrishnan S., Kapahi V.K., Rao A.P., Subrahmanya C.R., Kulkarni V.K., 1991, Current Science, 60, 95
\bibitem[\protect\citeauthoryear{Taylor et al. 1994}{}]{taylor94} 
Taylor C.L., Brinks E., Pogge R.W., Skillman E.D., 1994, AJ, 107, 971
\bibitem[\protect\citeauthoryear{Thuan, Izotov \& Foltz}{1999}]{thuan1999}
Thuan T.X., Izotov Y.I., Foltz C.B., 1999, ApJ, 525, 105
\bibitem[\protect\citeauthoryear{Tosi et al. 2006}{}]{tosi2006} 
Tosi M., Aloisi A., Mack J., Maio M., 2006, eprint (astro--ph/0609659)
\bibitem[\protect\citeauthoryear{Tully et al. 2008}{}]{Tully2007} 
Tully R.B., Shaya E.J., Karachentsev I.D., Courtois H.M., Kocevski D.D., Rizzi L., Peel A., 2008, ApJ, 676, 184
\bibitem[\protect\citeauthoryear{van Zee et al.}{1998}]{zee98IZw} 
van Zee L., Westpfahl D., Haynes M.P., Salzer J.J., 1998, AJ, 115, 1000
\bibitem[\protect\citeauthoryear{Zwicky}{1965}]{zwicky1965} 
Zwicky F., 1965, ApJ, 142, 1293
\end{thebibliography}
\end{document}